# Isotopic constraints on genetic relationships among group IIIF iron meteorites, Fitzwater Pass, and the Zinder pallasite.

Jonas Pape[1], Bidong Zhang[2], Fridolin Spitzer[3], Alan E. Rubin[2] and Thorsten Kleine[3]

[1]Institut für Planetologie, University of Münster, Wilhelm-Klemm-Str. 10,
48149 Münster, Germany

[2] Department of Earth, Planetary & Space Sciences, University of California, Los Angeles,
CA 90095-1567, USA

[3] Max Planck Institute for Solar System Research, Justus-von-Liebig-Weg 3,
37077 Göttingen, Germany

Corresponding author: Jonas Pape (jonas.pape@uni-muenster.de)

# Abstract

Complex interelement trends among magmatic IIIF iron meteorites are difficult to explain by fractional crystallization and have raised uncertainty about their genetic relationships. Nucleosynthetic Mo isotope anomalies provide a powerful tool to assess if individual IIIF irons are related to each other. However, while trace-element data are available for all nine IIIF irons, Mo isotopic data are limited to three samples. We present Mo isotopic data for all but one IIIF irons that help assess the genetic relationships among these irons, together with new Mo and W isotopic data for Fitzwater Pass (classified IIIF), and the Zinder pallasite (for which a cogenetic link with IIIF irons has been proposed). After correction for cosmic-ray exposure, the Mo isotopic compositions of the IIIF irons are identical within uncertainty and confirm their belonging to carbonaceous chondrite-type (CC) meteorites. The mean Mo isotopic composition of Group IIIF overlaps those Groups IIF and IID, but a common parent body for these groups is ruled out based on distinct trace element systematics. The new Mo isotopic data do not argue against a single parent body for the IIIF irons, and suggest a close genetic link among these samples. By contrast, Fitzwater Pass has distinct Mo and W isotopic compositions, identical to those of some non-magmatic IAB irons. The Mo and W isotope data for Zinder indicate that this meteorite is not related to IIIF irons, but belongs to the non-carbonaceous (NC) type and has the same Mo and W isotopic composition as main-group pallasites.

# Introduction

Most iron meteorites are thought to sample the metallic cores of early-formed differentiated planetesimals (Kruijer et al., 2014b; Anand et al., 2021), because their chemical compositions are consistent with variations expected from crystal-liquid fractionation imparted during crystallization of metallic magma (e.g., Scott 1972). These meteorites are referred to as "magmatic irons". Based on their chemical composition, the magmatic iron meteorites are classified into different chemical groups (e.g., Scott and Wasson, 1975), each of which is thought to represent the core of a single planetesimal (perhaps except for IIAB-IIG irons that may derive from the same asteroidal core; Wasson and Choe, 2009). In this context, group IIIF irons, which consists of only nine (or eight, as will be discussed later) members, is of special interest, because samples from this group do not seem to record consistent and coherent trace-element systematics. Hilton et al. (2022), based on highly siderophile element (HSE) data, argued that the chemical variations among the IIIF irons can be accounted for by simple fractional crystallization from a single metallic melt. However, Zhang et al. (2022) pointed out that the chemical variations among IIIF irons for some other elements (including in particular the Co, Ga, Ge, and Au) are difficult to account for by a simple fractional crystallization model.

Members of a given magmatic iron group should have identical nucleosynthetic isotope compositions, because they are presumed to have crystallized from a common metallic melt. Nucleosynthetic isotope anomalies, therefore, provide a powerful means for assessing as to whether the different members of a given iron meteorite group derive from a common parent body. These isotope anomalies arise through the heterogeneous distribution of presolar matter in the solar protoplanetary disk; they have been used to subdivide meteorites into non-carbonaceous (NC) and carbonaceous-chondrite (CC) types (Warren, 2011; Budde et al., 2016). For iron meteorites, differences in Mo isotopic composition are particularly useful in distinguishing these two types of meteorites, because CC-type materials are characterized by a systematic offset in Mo three-isotope plots caused by an excess in *r*-process Mo over NC-type materials (Budde et al., 2016; Poole et al., 2017; Kruijer et al., 2017) or a distinct *s*-process composition of the two reservoirs (Stephan and Davis, 2021). Based on the Mo isotopic composition of three IIIF irons - Klamath Falls, Clark County, and Oakley (iron) - it was concluded these samples formed in the CC reservoir. However, no Mo isotope data are available for any of the remaining six IIIF irons. Such data would help assess whether individual members of the IIIF group are genetically related and, hence, derive from a single parent body.

Beside the group-intern genetic relationships a potential co-genetic link between the IIIF irons and some pallasite meteorites has been proposed. The two pyroxene-bearing pallasites NWA 1911 and Zinder were suggested as being from the same parent asteroid as the IIIF irons, based mainly on chemical similarities between their metal phases (Boesenberg et al., 2012; Zhang et al, 2022). If the link is valid, these two pallasites and IIIF irons should share a common Mo isotopic composition.

We present new Mo isotope data for seven out of nine iron meteorites currently classified as IIIF and the pallasite Zinder. For Zinder and one of the IIIF irons (Fitzwater Pass) we also report W isotope data. For all samples, we also obtained Pt isotope data, to correct measured Mo and W isotope compositions for the effects of cosmic-ray exposure (CRE). These data are used to constrain the formation histories of these meteorites and to determine if they derive from a common body or multiple parent bodies.

## Materials and Methods

### Samples

Seven meteorites classified as group IIIF irons (Clark County, Nelson County, St. Genevieve County, Cerro del Inca, Oakley (iron), Moonbi, Fitzwater Pass) and the metal fraction of the pyroxene-bearing pallasite Zinder were selected for this study. Besides Clark County and Oakley (iron), none of these samples was previously studied for the isotope systematics of the siderophile elements Mo, W and Pt. Together with the iron meteorites Clark County, Klamath Falls, and Oakley (iron), for which isotope data have been published by Kruijer et al. (2017) and Worsham et al. (2019), Mo, W and Pt isotope data are now available for all but one meteorite (Binya) classified as IIIF.

### Analytical Methods

Iron meteorites were cut into 50–100 mg pieces using a diamond blade saw. Metal pieces of the pallasite Zinder were separated from the silicate phases using a pre-cleaned agate mortar dedicated for meteorite samples only. The metal chips of all samples were polished using SiC paper and sonicated in ethanol to remove saw marks and surface contamination. The samples were then digested in double-distilled 6M HCl + traces of conc. $HNO_3$ and aliquots for Mo-, W and Pt purification by ion exchange chromatography were taken from single digestion solutions. All purification procedures followed well-established protocols from our laboratory and are briefly described below.

The chemical separation of W and Mo followed the methods described by Kruijer et al. (2017) and Budde et al. (2018), which were adopted from Burkhardt et al. (2011, 2014). The digestion aliquots for W and Mo were converted to 0.5M HCl–0.5M HF and processed through first-stage anion-exchange columns filled with 4 ml AG1-X8 resin, where W and Mo were separated from the matrix, and eluted in 6M HCl–1M HF (W) and 3M $HNO_3$ (Mo). After repeated high-temperature dry-downs of the samples in concentrated HF, the W cuts were redissolved in 0.6M HF–0.4% $H_2O_2$ and loaded onto a second anion-exchange column. The final W cuts were eluted in 6M HCl–1M HF, repeatedly dried down in concentrated HF and converted into a 0.56M $HNO_3$–0.24M HF running solution for W isotope measurements. The Mo cuts collected during the W chemistry were further purified by a two-step ion-exchange chromatographic procedure (Budde et al., 2018). The Mo cuts were converted to 1M HCl, loaded onto columns filled with Eichrom TRU resin and Mo was eluted in 0.1M HCl. The samples were then dissolved in 7M $HNO_3$ and again loaded on columns filled with Eichrom TRU resin, where Mo was eluted in 0.1M $HNO_3$. The final Mo cuts were converted to a 0.5M $HNO_3$–0.01M HF running solution for Mo isotope analyses by MC-ICP-MS. Platinum was purified using a single-step anion-exchange column method developed by Rehkämper and Halliday (1997) and modified by Kruijer et al. (2013). In brief, the samples were converted to 1M HCl–10% $Br_2$ solution and loaded onto Quartz glass columns and the Pt cuts were eluted in 14 ml conc. $HNO_3$. After repeated high-temperature dry downs in conc. $HClO_4$ and subsequently in conc. $HNO_3$ the samples were finally dissolved in 0.5M $HNO_3$ running solution for MC-ICP-MS analysis. Typical yields of the chemical separation were ~60–90% for Pt and ~60–80% for W and Mo. The total procedural blanks were negligible for all elements.

All isotope measurements were performed on a Thermo Scientific *Neptune Plus* multicollector ICP-MS at the Institut für Planetologie in Münster following established analytical protocols from our laboratory. The samples were measured at concentrations of 50, 100 and 200 ppb for

W, Mo and Pt, respectively, and bracketed by measurements of terrestrial standard solutions. The isotope data are internally normalized using the exponential law and are reported as ε-unit (parts per $10^4$) deviation from terrestrial standard values. For N≥4 measurements of the same sample, uncertainties are reported as 95% confidence intervals of the mean. For N<4 the uncertainties are given as the external reproducibility (2 s.d.) of the reference material NIST SRM 129c or the internal error (2 s.e.) of a single measurement, whichever is larger.

The precision and accuracy of the Mo, W, and Pt isotope measurements were assessed through multiple measurements of the terrestrial reference material NIST SRM 129c (doped with Pt) that was processed together with the meteorite samples. The Mo, W, and Pt isotope data of the NIST129c measurements are reported in the Supplementary Material (Tables S1–S3) and are in good agreement with data from previous studies. Tungsten isotope data of NIST SRM 129c involving $^{183}$W show a small mass-independent effect resulting in small excesses for ε$^{182}$W (normalized to $^{186}$W/$^{183}$W or "6/3") and ε$^{184}$W ("6/3") and corresponding deficits for ε$^{183}$W ("6/4"). This effect has been observed in several previous studies and is likely introduced during the W purification procedure (Willbold et al., 2011; Kruijer et al., 2017; Budde et al., 2022). The magnitude of this analytical $^{183}$W effect can vary between different laboratories and studies and commonly results in a deficit of <0.1 ε-units on ε$^{183}$W ("6/4"). All W isotope data involving $^{183}$W reported in this study as *measured* values are corrected for this effect using the magnitude determined by repeated measurements of the NIST SRM 129c reference material in this study and following the method described in Kruijer et al. (2017).

# Results

**Pt isotopes and correction for cosmic-ray exposure**
The Mo and W isotopic compositions of iron meteorites can be affected by secondary neutron capture processes due to cosmic-ray exposure (CRE). These CRE effects can be corrected using Pt isotopes as a neutron dosimeter (Kruijer et al., 2013; Wittig et al., 2013; Spitzer et al., 2020). Because CRE effects vary with depth, the Pt, Mo, and W isotopic compositions should ideally be acquired from the same meteorite chips. For this study, the Mo, W and Pt isotope data have been obtained from aliquots of the same sample solution for all but one sample. Only for Oakley (iron) we used the Pt isotope data from a previous study (Worsham et al., 2019), but the Mo isotopic data of this study were obtained from a remaining metal piece from the previous study and, thus, essentially from the same samples. This is confirmed by the identical Mo isotope composition determined for the two different sample solutions in the two studies (see below).

The Pt isotope data of all analyzed IIIF meteorites, Fitzwater Pass, and Zinder are listed in Table 1. Cosmic-ray exposure leads to an upward shift of $\varepsilon^{196}$Pt (normalized to $^{198}$Pt/$^{195}$Pt, "8/5"), meaning that more positive $\varepsilon^{196}$Pt (8/5) values indicate stronger CRE effects. The $\varepsilon^{196}$Pt (8/5) values for the samples of this study range from almost unirradiated (0.03±0.08, Nelson County) to considerably irradiated (0.48±0.06, Oakley). This range is similar to the range observed for other iron meteorite groups (e.g., Hunt et al., 2017; Kruijer et al, 2017). The Zinder pallasite records only weak irradiation with $\varepsilon^{196}$Pt = 0.07±0.04.

Because nucleosynthetic anomalies are generally small for $\varepsilon^{196}$Pt (8/5) (see below and Spitzer et al., 2021), correlations of $\varepsilon^{196}$Pt (8/5) versus $\varepsilon^{i}$Mo or $\varepsilon^{182}$W can be used to correct measured Mo and W isotopic compositions for CRE-induced neutron capture (Fig. 1). For W isotopes, a well-defined weighted mean regression for $\varepsilon^{196}$Pt (8/5) versus $\varepsilon^{182}$W (6/4) of −1.320±0.055 has been determined based on seven magmatic iron-meteorite groups (from both the NC and CC reservoirs) by Kruijer et al. (2017) and this slope has been used to correct the W isotope data of Fitzwater Pass and Zinder from this study. For Mo isotopes, the new data from this study sampling almost all group IIIF irons make it possible to determine precise group-internal $\varepsilon^{196}$Pt (8/5) versus $\varepsilon^{i}$Mo regressions (Fig. 1). The slopes of these regressions are in good agreement with those for the IC, IID, IIAB, and IIIAB iron-meteorite groups reported in Spitzer et al. (2020b) (Fig. S1). The weighted mean slopes for all five groups are: −0.461±0.069 for $\varepsilon^{92}$Mo, −0.3±0.058 for $\varepsilon^{94}$Mo, −0.371±0.042 for $\varepsilon^{95}$Mo, −0.092±0.078 for $\varepsilon^{97}$Mo, and 0.130±0.057 for $\varepsilon^{100}$Mo. These slopes were used for the correction of CRE-effects on Mo isotopes for the individual IIIF irons in this study (Fig. 2). Finally, Spitzer et al. (2021) recently showed that there are small nucleosynthetic Pt isotope anomalies in iron meteorites and that an $\varepsilon^{196}$Pt value of −0.06 is the most appropriate value for the pre-exposure Pt isotopic composition of iron meteorites; this value was used for all CRE corrections in this study.

## Mo Isotopes

The Mo isotope data are listed in Table 2 and are shown in Figures 2 and 3. The measured (i.e., CRE-uncorrected) Mo isotope compositions of Clark County and Oakley (iron) agree well with previously published values for the same samples (Kruijer et al., 2017; Worsham et al., 2019). The CRE-corrected Mo isotope data are shown in Figure 3 along with published Mo isotope data for other iron-meteorite groups. All analyzed IIIF irons but Fitzwater Pass have well-resolved $\varepsilon^i$Mo nucleosynthetic anomalies and have the same CRE-corrected Mo isotopic composition within uncertainties. In the $\varepsilon^{94}$Mo vs. $\varepsilon^{95}$Mo diagram (Fig. 3), they plot on the CC-line close to the IIF and IID iron groups. Fitzwater Pass, currently classified as a IIIF iron in the Meteoritical Bulletin Database, displays a Mo isotopic composition clearly different from the IIIF irons and plots on the NC-line close to the origin. The non-magmatic IAB irons are the only iron meteorites known to have a similar Mo isotopic composition (Poole et al., 2017; Worsham et al., 2017). This is in line with recent trace-element analyses of Fitzwater Pass, which also suggest a link between Fitzwater Pass and group IAB irons (Zhang et al., 2022). Finally, the Mo isotope composition of the metal fraction of Zinder, for which a genetic link to the IIIFs has been proposed (Zhang et al., 2022), is different from the IIIF irons and plots on the NC-line in the $\varepsilon^{94}$Mo–$\varepsilon^{95}$Mo diagram.

## W Isotopes

The W isotope data for Fitzwater Pass and Zinder are listed in Table 3 and are shown in Figure 4. After correction for a small mass-independent effect, both samples show no resolved $^{183}$W anomaly with $\varepsilon^{183}$W (6/4) values of 0.01±0.03 for Fitzwater Pass and 0.00±0.05 for Zinder. These values agree with those of NC iron meteorites, whereas irons from the CC reservoir typically show small positive $\varepsilon^{183}$W values (Qin et al., 2008; Kruijer et al., 2013; Kruijer et al., 2017). Measured $\varepsilon^{182}$W (6/4) values were first corrected for nucleosynthetic effects using the $\varepsilon^{183}$W (6/4) of the same sample and then corrected for CRE effects as described above. The final pre-exposure $\varepsilon^{182}$W (6/4) of Zinder is −3.17±0.10, which overlaps with the published mean value for the IIIF irons Clark County and Klamath Falls of −3.24±0.10 (Kruijer et al., 2017). However, the pre-exposure $\varepsilon^{182}$W (6/4) for Fitzwater Pass of −2.69±0.08 is more radiogenic than any other pre-exposure $\varepsilon^{182}$W of magmatic iron meteorites, but agrees with pre-exposure $\varepsilon^{182}$W values reported for some non-magmatic IIE irons (e.g., −2.63±0.12,

Weekeroo Station; Kruijer and Kleine, 2019) and is similar to some IAB irons (e.g., −2.88±0.06, Persimmon Creek; Worsham et al., 2017).

## Discussion

### Mo isotope evidence for a common parent body of IIIF irons

The Mo isotopic compositions of the seven IIIF irons and Zinder obtained in this study together with previously published Mo data for Klamath Falls (Kruijer et al., 2017) can be used to assess potential genetic links among these samples. To this end, the $\varepsilon^{94}$Mo–$\varepsilon^{95}$Mo systematics are particularly suited, because they can be used to track the genetic heritage of meteorites from either the NC or CC reservoir (e.g., Budde et al., 2016) and also of material with mixed NC-CC heritage (Spitzer et al., 2022). Additionally, this system allows the resolution of several but not all iron meteorite groups within the CC reservoir. We will first focus on the Mo isotope systematics of seven of the IIIF irons (Clark County, Nelson County, St. Genevieve County, Cerro del Inca, Oakley (iron), Moonbi, Klamath Falls). The results for Fitzwater Pass and the Zinder pallasite are discussed separately, because these two samples have distinct Mo isotopic compositions. In this section, we accordingly do not treat Fitzwater Pass as a IIIF iron.

A key finding of this study is that all analysed IIIF samples have very similar Mo isotope compositions, and when corrected for CRE-effects show no resolved Mo isotope variability beyond analytical uncertainty (Fig. 3). Thus, all IIIF irons for which Mo isotope data are now available (i.e., all except Binya) have the same Mo isotopic composition. With these data, precise group IIIF mean values are calculated either from the intercept of the $\varepsilon^{196}$Pt–$\varepsilon^{i}$Mo slopes or as the weighted mean of the individual CRE-corrected samples, which both yield the same results (Table 2). The Mo isotope data, therefore, are consistent with a common genetic heritage of the different IIIF samples. This is also consistent with the elemental HSE systematics of the IIIFs, which can be successfully modeled by simple fractional crystallization from a single metallic melt (Hilton et al., 2022). Nevertheless, Zhang et al. (2022) have shown that this is not true for other elements such as Co, Ga, Ge, and Au, and on this basis questioned a common genetic heritage of all IIIF irons. To this end, the new Mo data can be evaluated within the framework of three different scenarios: *i*) individual or several IIIF irons do not belong to the IIIF group but belong instead to other CC iron groups, *ii*) some IIIF irons should be discarded from the IIIF group and be reclassified as *ungrouped* irons, and *iii*) all IIIF irons share a common heritage and thus form a coherent group.

In scenario *i*) the IIF and IID irons are the only groups that are isotopically close to the IIIF irons. Group IIF is, similar to IIIF, a small group and Mo isotope data are available for only two samples (Kruijer et al., 2017; Worsham et al., 2019), while the IID group consists of more samples and Mo isotope data have been reported for four members (Burkhardt et al., 2011; Kruijer et al., 2017; Worsham et al., 2019, Spitzer et al., 2020b). The IIF, IID, and IIIF groups all have indistinguishable Mo isotope compositions (Fig. 3), so that a clear distinction of samples from these three groups is not possible based on Mo isotopes alone. However, a genetic heritage of any of the IIIF samples to either of the two other groups can be ruled out by also considering their trace-element abundances. In particular, Ga/Ni and Ge/Ni ratios for all IIIF iron meteorites are significantly lower than those of IIF and IID irons, indicating that the IIIF irons are more strongly depleted in volatile elements (Fig. S2). It can, therefore, be ruled out that any of the IIIF irons belongs to either the IIF or IID group.

In scenario *ii*) it is important to recognize that from a trace-element perspective, several samples appear to not fit the expected fractionation trends of a common magmatic IIIF group (Zhang et al., 2022). As such, exclusion of single samples from the IIIF group would not solve the issue that members of this group do not define consistent fractional-crystallization trends. Instead, all IIIF irons plot in very narrow fields in Ga vs. Ni and Ge vs. Ni space, and there are almost no ungrouped irons known that plot close to the IIIF compositional field (Goldstein et al., 2009). Together with the homogeneous Mo isotopic composition of all IIIF irons, these observations make it highly unlikely that the IIIF label applies to individual *ungrouped* irons rather than to genetically linked samples.

This leaves option *iii*), a common heritage of Group IIIF irons and thus a common parent body for these irons as the most likely interpretation of the combined Mo isotope and trace-element data. This does not exclude the possibility that the IIIF group is subdivided into two grouplets with distinct chemical-fractionation histories, namely the Moonbi grouplet (Moonbi, St. Genevieve County, Cerro del Inca) and the Clark County grouplet (Clark County, Nelson County, Oakley) (Zhang et al., 2022). However, the Mo isotope data and Ge-Ga-Ni systematics would then require that these two grouplets are genetically related, which again permits combining these samples into a single IIIF group.

**A non-magmatic origin of Fitzwater Pass**

Fitzwater Pass was originally classified as one of nine IIIF iron meteorites and is (July 2023) still listed as IIIF in the Meteoritical Bulletin Database. However, new LA-ICP-MS trace-element data for this sample (Zhang et al., 2022) have shown that Fitzwater Pass has significantly higher Ge and Ga concentrations than other IIIF irons (as reported in its original classification), which makes Fitzwater Pass chemically more similar to IAB than IIIF. Consistent with this, the Mo isotopic composition of Fitzwater Pass (with $\varepsilon^{94}$Mo of 0.14±0.09 and $\varepsilon^{95}$Mo of 0.02±0.04) demonstrates that this iron is not of CC origin (in contrast to all other IIIF irons), but instead plots on the NC-line, where it overlaps with the composition of some IAB irons (Fig. 3). Moreover, although some IAB irons, especially those from the sHH and sHL subgroups, have larger Mo nucleosynthetic isotope anomalies than most other IAB samples (Worsham et al., 2017), there is no other iron-meteorite group with a Mo isotope composition similar to that of Fitzwater Pass. As such, the new Mo isotope data, together with the trace-element data, suggest that Fitzwater Pass is genetically related to group IAB irons. Reclassification of Fitzwater Pass as a IAB iron meteorite, however, will require more comprehensive trace element modeling.

The pre-exposure $^{182}$W isotope composition of Fitzwater Pass determined in this study provides further evidence for a non-magmatic origin of this sample. The $^{182}$W data can be used to calculate a Hf-W model age for metal-silicate equilibration using the following equation:

$$\Delta t_{CAI} = -\frac{1}{\lambda} ln \left[ \frac{(\varepsilon^{182}W)_{sample} - (\varepsilon^{182}W)_{chondrites}}{(\varepsilon^{182}W)_{SSI} - (\varepsilon^{182}W)_{chondrites}} \right] \quad (1)$$

where $\lambda$ is the $^{182}$Hf decay constant of 0.0778 ± 0.0015 Ma$^{-1}$, $(\varepsilon^{182}W)_{chondrites}$ is the present-day composition of carbonaceous chondrites of −1.91 ± 0.08 (Kleine et al., 2004; 2009), and $(\varepsilon^{182}W)_{SSI}$ is the solar system initial of −3.49 ± 0.07 obtained from calcium-aluminum-rich inclusions (CAIs) (Kruijer et al., 2014a). This equation assumes a single-stage metal-silicate separation (i.e., Hf-W fractionation) from a chondritic Hf/W reservoir. The pre-exposure $^{182}$W of Fitzwater Pass of −2.69±0.08 is significantly higher than values observed for magmatic iron meteorites and, accordingly, translates into a relatively young Hf-W model age of 9.1 ± 1.7 Ma after CAIs. Among iron meteorites, such radiogenic pre-exposure $\varepsilon^{182}$W values are only known from the non-magmatic groups IAB (Worsham et al., 2017) and IIE (Kruijer and Kleine, 2019); as such, the $^{182}$W data suggest a non-magmatic origin of Fitzwater Pass and support a potential genetic link of this sample to the IAB group.

**Zinder is not related to IIIF irons**

Because pallasites can mineralogically be viewed as a mixture of iron meteorites and ultramafic stony meteorites, finding a genetic link of any pallasite meteorite to either metal or silicates from other meteorites holds important clues for understanding the formation history of pallasites (e.g., Greenwood et al., 2015; Kruijer et al., 2022; Windmill et al., 2022). In this context, based on chemical similarities of metal in the pyroxene-bearing Zinder pallasite with some IIIF irons, it was proposed that Zinder might be genetically related to group IIIF (Boesenberg et al., 2012; Zhang et al., 2022). The O isotope composition of Zinder rather points towards an NC origin of Zinder, however (Humayun et al., 2018). The new Mo isotopic data of this study demonstrate that Zinder's metal phase shows no affinity to IIIF irons, but instead derives from the NC reservoir, consistent with its O isotope composition. Additionally, Zinder has the same Mo isotopic composition as reported recently for main-group pallasites (Kruijer et al., 2022). This is further supported by W isotope data for which both the $\varepsilon^{183}$W of Zinder (0.00±0.05) as well as the pre-exposure $\varepsilon^{182}$W (–3.17±0.10) are indistinguishable from those of main-group pallasites and different from any CC iron (Fig. 4). Together, these new data effectively rule out a genetic link of Zinder and group IIIF irons; instead, these data would suggest a link to main-group pallasites. However, in O isotope space, Zinder plots almost on the terrestrial fractionation line (TFL) ($\Delta^{17}$O = +0.05 ‰; Bunch et al., 2005), while the main-group pallasite's average $\Delta^{17}$O is −0.187±0.016‰ (Greenwood et al., 2017). These data suggest that Zinder, despite its NC origin, is also not directly related to main-group pallasites.

# Conclusions

The Mo isotopic compositions of the seven iron meteorites analyzed in this study place new constraints on the genetic relations among IIIF iron meteorites. All seven irons have the same Mo isotopic composition (agreeing with the earlier published Mo isotope composition of Klamath Falls); this strongly suggests a genetic link between all samples currently classified as IIIF (with the exception of Fitzwater Pass). The Mo isotope composition of group IIIF is similar to that of the IIF or IID groups, but the distinct chemical composition of the IIIFs (especially Ga and Ge) indicate that the IIIF irons are not directly linked to the IIF or IID irons. The identical Mo isotopic composition of the IIIF irons together with their unique Ga/Ni and Ge/Ni ratios, therefore, indicate that IIIF irons define a coherent chemical group and that all IIIF irons are from a common parent body. As such, the difficulty to account for chemical variations among the IIIF irons by fractional crystallization of a single metallic magma implies that the IIIF core underwent more complex chemical processes, the nature of which remains to be identified.

The Mo and W isotopic data of Fitzwater Pass show that this iron does not belong to the IIIF group, but is instead from the NC reservoir and may be genetically related to the IAB irons. However, whether or not Fitzwater Pass should be re-classified as IAB will require additional work.

Finally, the Mo isotope systematics of the pyroxene-bearing pallasite Zinder indicate that this sample is not related to the IIIF irons. Instead, Zinder's Mo isotopic composition indicates derivation from the NC reservoir and overlaps with the Mo isotope composition of main-group pallasites. However, differences in O isotope composition between Zinder and main-group pallasites seem to rule out a direct genetic link between these pallasites.

# Acknowledgements

We are honored to contribute to this special issue in memoriam of Edward Robert Dalton Scott, who made so many fundamental contributions to the field of meteoritics, and in particular to our understanding of the origin of iron meteorites and the crucial role they play in unraveling the earliest history of the Solar System. We thank Richard Walker and associate editor Sasha Krot for constructive reviews and editorial handling of this manuscript. Funded by the Deutsche Forschungsgemeinschaft (DFG, German Research Foundation) – Project-ID 263649064 – TRR 170. This research was also supported by NASA grants NNX17AE77G (AER) and 0NSSC19K1238 and 80NSSC23K0035 (BZ). This is TRR 170 publication no. 201.

# References

Anand A., Pape J., Wille M., Mezger K., and Hofmann B. 2021. Early differentiation of magmatic iron meteorite parent bodies from Mn-Cr chronometry. *Geochemical Perspective Letters* 20:6–10.

Bermingham K. R., Worsham E. A., and Walker R. J. 2018. New insights into Mo and Ru isotope variation in the nebula and terrestrial planet accretionary genetics. *Earth and planetary science letters* 487:221–229.

Boesenberg J. S., Delaney J. S., and Hewins R. H. 2012. A petrological and chemical reexamination of Main Group pallasite formation. *Geochimica et Cosmochimica Acta* 89:134–158.

Budde G., Burkhardt C., Brennecka G. A., Fischer-Gödde M., Kruijer T. S., and Kleine T. 2016. Molybdenum isotopic evidence for the origin of chondrules and a distinct genetic heritage of carbonaceous and noncarbonaceous meteorites. *Earth and Planetary Science Letters* 454:293–303.

Budde G., Kruijer T. S., and Kleine T. 2018. Hf-W chronology of CR chondrites: Implications for the timescales of chondrule formation and the distribution of $^{26}$Al in the solar nebula. *Geochimica et Cosmochimica Acta* 222:284–304.

Budde G., Archer G. J., Tissot F. L. H., Trappe S., and Kleine T. 2022. Origin of the analytical $^{183}$W effect and its implications for tungsten isotope analyses. *Journal of Analytical Atomic Spectrometry* 37:2005–2021.

Bunch T. E., Rumble III D., Wittke J. H., and Irving A. J. 2005. Pyroxene-rich pallasites Zinder and NWA 1911: not like the others. *Meteoritics and Planetary Science Supplement* 40, 5219.

Burkhardt C., Kleine T., Oberli F., Pack A., Bourdon B., and Wieler R. 2011. Molybdenum isotope anomalies in meteorites: Constraints on solar nebula evolution and origin of the Earth. *Earth and Planetary Science Letters* 312:390–400.

Burkhardt C., Hin R. C., Kleine T., and Bourdon B. 2014. Evidence for Mo isotope fractionation in the solar nebula and during planetary differentiation. *Earth and Planetary Science Letters* 391:201–211.

Goldstein J. I., Scott E. R. D., and Chabot N. L. 2009. Iron meteorites: Crystallization, thermal history, parent bodies, and origin. *Geochemistry* 69:293–325.

Greenwood R. C., Barrat J. A., Scott E. R., Haack H., Buchanan P. C., Franchi I. A., Yamaguchi A., Johnson D., Bevan A. W. R., and Burbine, T. H. 2015. Geochemistry and oxygen isotope composition of main-group pallasites and olivine-rich clasts in mesosiderites: Implications for the "Great Dunite Shortage" and HED-mesosiderite connection. *Geochimica et Cosmochimica Acta* 169:115–136.

Greenwood R. C., Burbine T. H., Miller M. F., and Franchi I. A. 2017. Melting and differentiation of early-formed asteroids: The perspective from high precision oxygen isotope studies. *Geochemistry* 77, 1–43.

Hilton C. D., Ash R. D. and Walker R. J. 2022. Chemical characteristics of iron meteorite parent bodies. *Geochimica et Cosmochimica Acta* 318, 112–125.

Humayun M., Boesenberg J. S., and van Niekirk D. 2018. Composition of the IIIF irons and their relationship to the Zinder pallasite. In In *49th Lunar and planetary science conference*, p. 1461.

Hunt A. C., Ek M., and Schönbächler M. 2017. Platinum isotopes in iron meteorites: Galactic cosmic ray effects and nucleosynthetic homogeneity in the p-process isotope $^{190}$Pt and the other platinum isotopes. *Geochimica et Cosmochimica Acta* 216:82–95.

Kleine T., Mezger K., Münker C., Palme H., and Bischoff A. 2004. $^{182}$Hf-$^{182}$W isotope systematics of chondrites, eucrites, and martian meteorites: Chronology of core formation and early mantle differentiation in Vesta and Mars. *Geochimica et Cosmochimica Acta* 68:2935–2946.

Kleine T., Touboul M., Bourdon B., Nimmo F., Mezger K., Palme H., Jacobsen S. B., Yin Q.-Z., and Halliday A. N. 2009. Hf-W chronology of the accretion and early evolution of asteroids and terrestrial planets. *Geochimica et Cosmochimica Acta* 73:5150–5188.

Kruijer T. S., Fischer-Gödde M., Kleine T., Sprung P., Leya I., and Wieler R. 2013. Neutron capture on Pt isotopes in iron meteorites and the Hf-W chronology of core formation in planetesimals. *Earth and Planetary Science Letters* 361:162–172.

Kruijer T. S., Kleine T., Fischer-Gödde M., Burkhardt C., and Wieler R. 2014a. Nucleosynthetic W isotope anomalies and the Hf-W chronometry of Ca-Al-rich inclusions. *Earth and Planetary Science Letters* 403:317327.

Kruijer T. S., Touboul M., Fischer-Gödde M., Bermingham K. R., Walker R. J., and Kleine T. 2014b. Protracted core formation and rapid accretion of protoplanets. *Science* 344:1150–1154.

Kruijer T. S., Burkhardt C., Budde G., and Kleine T. 2017. Age of Jupiter inferred from the distinct genetics and formation times of meteorites. *Proceedings of the National Academy of* Sciences 114:6712–6716.

Kruijer T. S. and Kleine T. 2019. Age and origin of IIE iron meteorites inferred from Hf-W chronology. *Geochimica et Cosmochimica Acta* 262:92–103.

Kruijer T. S., Burkhardt C., Borg L. E., and Kleine T. 2022. Tungsten and molybdenum isotopic evidence for an impact origin of pallasites. *Earth and Planetary Science Letters* 584:117440.

Poole G. M., Rehkämper M., Coles B. J. Goldberg, T., and Smith C. L. 2017. Nucleosynthetic molybdenum isotope anomalies in iron meteorites–new evidence for thermal processing of solar nebula material. *Earth and Planetary Science Letters*, 473:215–226.

Qin L., Dauphas N., Wadhwa M., Masarik J., and Janney, P. E. 2008. Rapid accretion and differentiation of iron meteorite parent bodies inferred from $^{182}$Hf–$^{182}$W chronometry and thermal modeling. *Earth and Planetary Science Letters* 273:94–104.

Rehkämper M. and Halliday A. N. 1997. Development and application of new ion-exchange techniques for the separation of the platinum group and other siderophile elements from geological samples. *Talanta* 44:663–672.

Scott E. D. 1972. Chemical fractionation in iron meteorites and its interpretation. *Geochimica et Cosmochimica Acta* 36:1205–1236.

Scott E. R. D. and Wasson J. T. 1975. Classification and properties of iron meteorites. *Reviews of Geophysics* 13:527–546.

Spitzer F., Burkhardt C., Budde G., Kruijer T., Morbidelli A., and Kleine T. 2020b. Isotopic evolution of the inner solar system inferred from molybdenum isotopes in meteorites. *The Astrophysical Journal* 898:L2.

Spitzer F., Burkhardt C., Nimmo F., and Kleine T. 2021. Nucleosynthetic Pt isotope anomalies and the Hf-W chronology of core formation in inner and outer solar system planetesimals. *Earth and Planetary Science Letters* 576:117211.

Spitzer F., Burkhardt C., Pape J., and Kleine T. 2022. Collisional mixing between inner and outer solar system planetesimals inferred from the Nedagolla iron meteorite. *Meteoritics & Planetary Science* 57:261–276.

Stephan T. and Davis A. M. 2021. Molybdenum isotope dichotomy in meteorites caused by s-process variability. *The Astrophysical Journal* 909(1), 8.

Warren P. H. 2011. Stable-isotopic anomalies and the accretionary assemblage of the Earth and Mars: A subordinate role for carbonaceous chondrites. *Earth and Planetary Science Letters* 311:93–100.

Wasson J. T. and Choe W.-H. 2009. The IIG iron meteorites: Probable formation in the IIAB core. *Geochimica et Cosmochimica Acta* 73:4879-4890.

Willbold M., Elliott T., and Moorbath S. 2011. The tungsten isotopic composition of the Earth's mantle before the terminal bombardment. *Nature* 477:195–198.

Windmill R. J., Franchi I. A., Hellmann J. L., Schneider J. M., Spitzer F., Kleine T., Greenwood R., and Anand M. 2022. Isotopic evidence for pallasite formation by impact mixing of olivine and metal during the first 10 million years of the Solar System. *PNAS Nexus*, *1*(1), pgac015.

Wittig N., Humayun M., Brandon A. D., Huang S., and Leya, I. 2013. Coupled W–Os–Pt isotope systematics in IVB iron meteorites: In situ neutron dosimetry for W isotope chronology. *Earth and Planetary Science Letters* 361:152–161.

Worsham E. A., Bermingham K. R., and Walker R. J. 2017. Characterizing cosmochemical materials with genetic affinities to the Earth: Genetic and chronological diversity within the IAB iron meteorite complex. *Earth and Planetary Science Letters* 467:157–166.

Worsham E. A., Burkhardt C., Budde G., Fischer-Gödde M., Kruijer T. S., and Kleine T. 2019. Distinct evolution of the carbonaceous and non-carbonaceous reservoirs: Insights from Ru, Mo, and W isotopes. *Earth and Planetary Science Letters* 521:103–112.

Yokoyama T., Nagai Y., Fukai R., and Hirata T. 2019. Origin and evolution of distinct molybdenum isotopic variabilities within carbonaceous and noncarbonaceous reservoirs. *The Astrophysical Journal* 883:62.

Zhang B., Chabot N. L., Rubin A. E., Humayun M., Boesenberg J. S., and van Niekerk D. 2022. Chemical study of group IIIF iron meteorites and the potentially related pallasites Zinder and Northwest Africa 1911. *Geochimica et Cosmochimica Acta* 323:202–219.

**Figures**

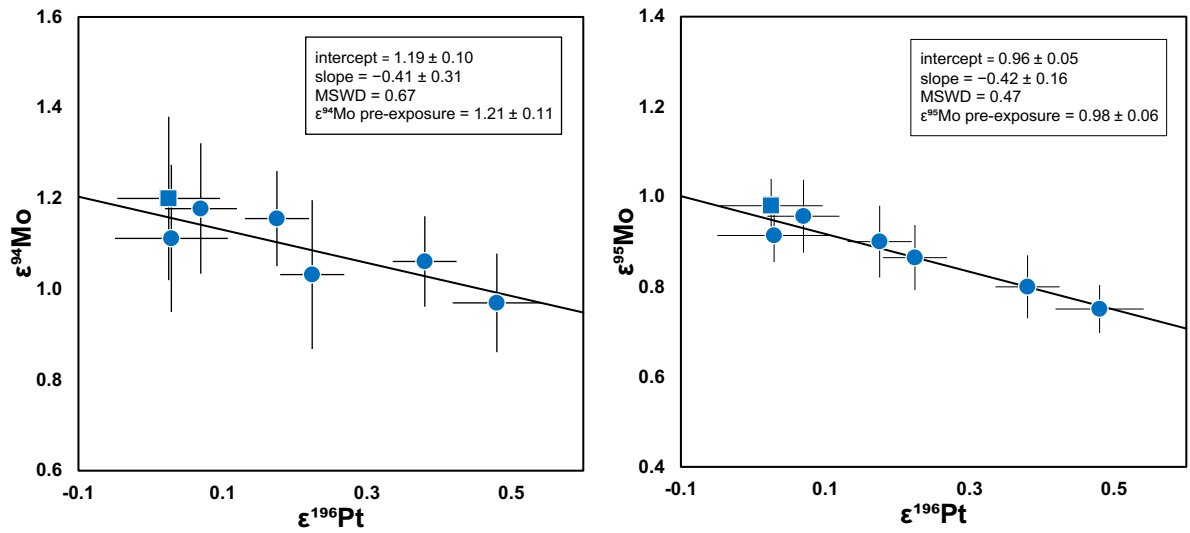

Fig. 1: $\varepsilon^{94}$Mo vs. $\varepsilon^{196}$Pt and $\varepsilon^{95}$Mo vs. $\varepsilon^{196}$Pt slopes for IIIF iron meteorites analyzed in this study (circles) and Klamath Falls (square; Kruijer et al., 2017) showing the negative correlation between the $\varepsilon^{94, 95}$Mo anomaly and the magnitude of cosmic-ray exposure.

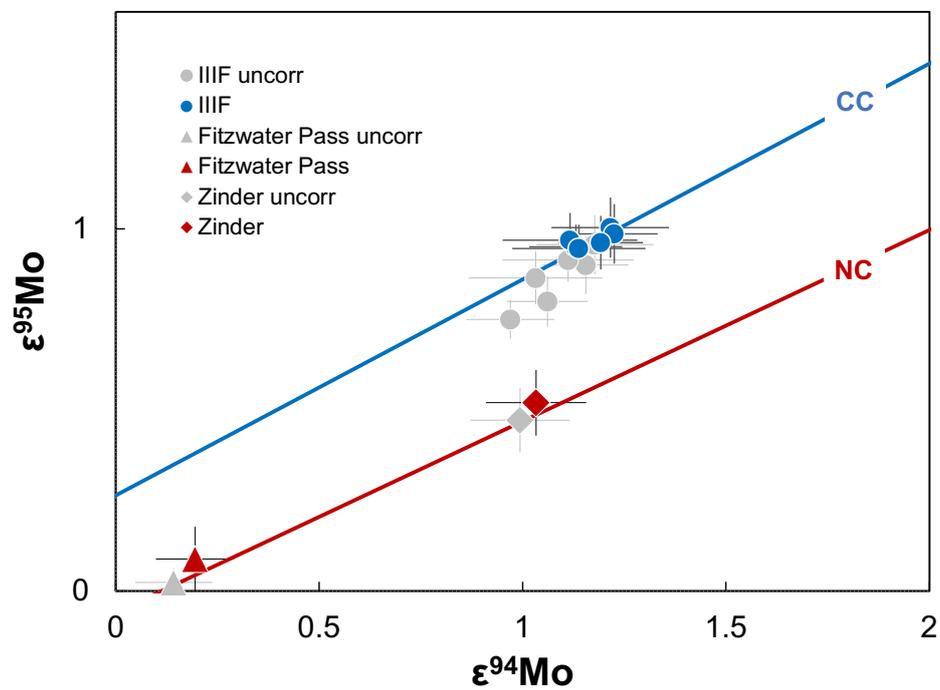

Fig. 2: Measured vs. CRE-corrected Mo isotope data for all analyzed IIIF irons, Fitzwater Pass and the Zinder pallasite.

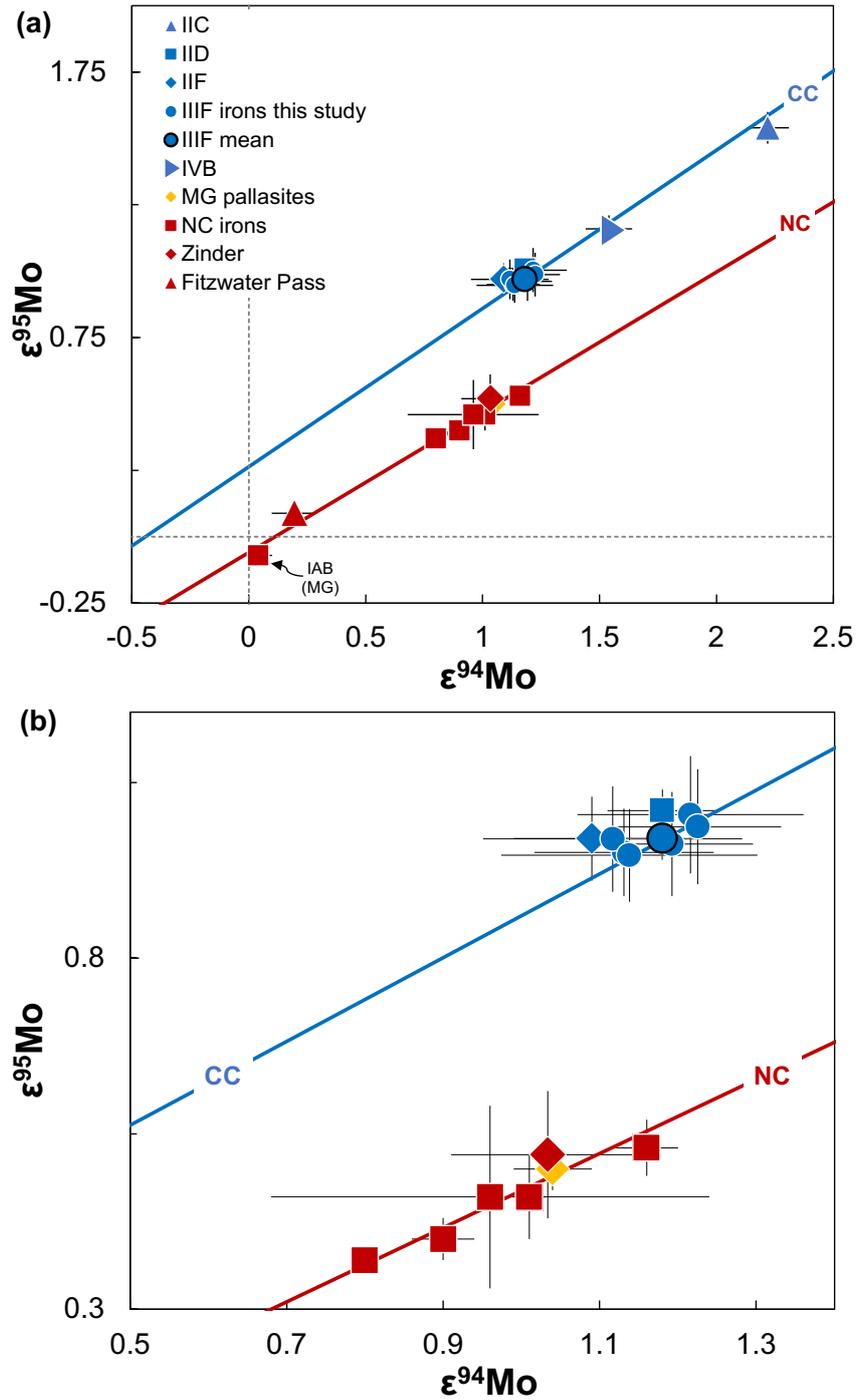

Fig. 3: Plots of $\varepsilon^{94}$Mo vs. $\varepsilon^{95}$Mo showing the IIIF Mo data along with literature data for other NC and CC iron meteorite groups (Bermingham et al., 2018; Budde et al., 2019; Worsham et al., 2019; Yokoyama et al., 2019; Spitzer et al., 2020b) and Mo data for main-group pallasites (Kruijer et al., 2022). Note that some IAB irons record $\varepsilon^{94,\,95}$Mo anomalies of up to ~1.2 and ~0.5, respectively (Worsham et al., 2019).

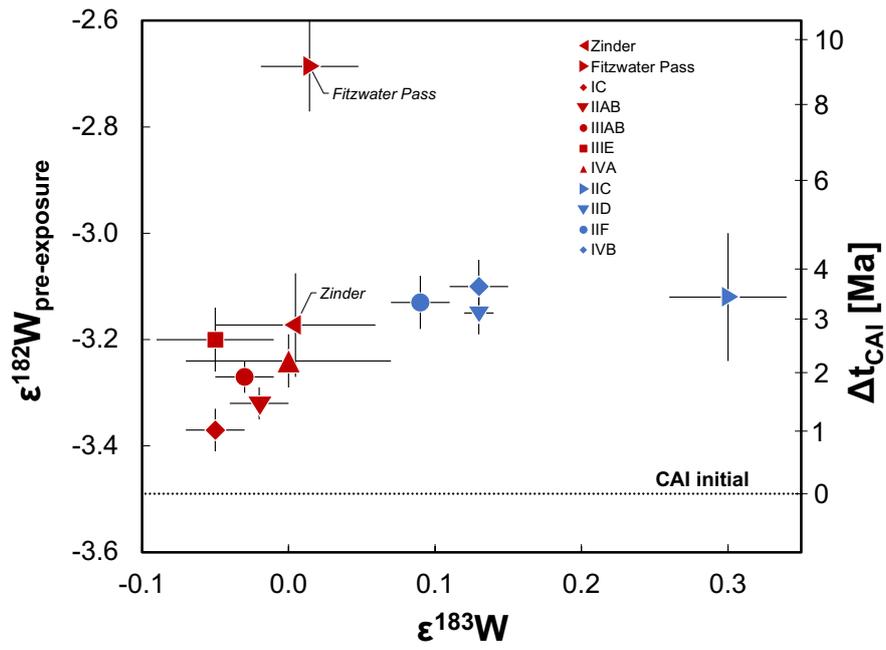

Fig. 4: W isotope data of iron meteorite Fitzwater Pass and the Zinder pallasite along with data for magmatic iron meteorites from the NC and CC reservoirs (Kruijer et al., 2017). Note that the $\varepsilon^{182}W$ data taken from Kruijer et al. (2017) were recalculated taking into account a small nucleosynthetic $^{196}Pt$ anomaly (see Spitzer et al., 2021).

# Tables

TABLE 1. Pt isotope data for IIIF iron meteorites, Fitwater Pass and pyroxene-bearing pallasite Zinder.

| Meteorite | Collection ID | N | Normalized to $^{196}Pt/^{195}Pt$ | | | Normalized to $^{198}Pt/^{195}Pt$ | | |
|---|---|---|---|---|---|---|---|---|
| | | | $\varepsilon^{192}Pt$ (± 2σ) | $\varepsilon^{194}Pt$ (± 2σ) | $\varepsilon^{198}Pt$ (± 2σ) | $\varepsilon^{192}Pt$ (± 2σ) | $\varepsilon^{194}Pt$ (± 2σ) | $\varepsilon^{196}Pt$ (± 2σ) |
| **IIIF** | | | | | | | | |
| Cerro del Inca | UCLA 3092 | 3 | 4.50 ± 0.82 | 0.53 ± 0.11 | -0.67 ± 0.13 | 3.91 ± 0.65 | 0.29 ± 0.09 | 0.22 ± 0.04 |
| Clark County | BM 1959, 949 | 4 | 4.46 ± 0.54 | 0.27 ± 0.04 | -0.21 ± 0.15 | 4.29 ± 0.57 | 0.20 ± 0.03 | 0.07 ± 0.05 |
| Moonbi | UCLA S2203 | 3 | 2.46 ± 0.82 | 0.33 ± 0.11 | -0.52 ± 0.13 | 1.87 ± 0.65 | 0.15 ± 0.09 | 0.18 ± 0.04 |
| Nelson County | UCLA S2201 | 4 | 1.64 ± 0.76 | 0.07 ± 0.14 | -0.09 ± 0.23 | 1.56 ± 0.52 | 0.06 ± 0.05 | 0.03 ± 0.08 |
| St. Genevieve County | UCLA 737 | 3 | 7.32 ± 0.82 | 0.79 ± 0.11 | -1.13 ± 0.13 | 6.15 ± 0.65 | 0.40 ± 0.09 | 0.38 ± 0.04 |
| Klamath Falls [1] | ME 2789 | 2 | 0.98 ± 1.30 | 0.11 ± 0.13 | -0.08 ± 0.22 | 0.69 ± 1.30 | 0.10 ± 0.11 | 0.03 ± 0.07 |
| Oakley (iron) [2] | USNM 780 | 4 | | | | 18.93 ± 3.33 | 0.78 ± 0.09 | 0.48 ± 0.06 |
| Fitzwater Pass | CML 0413-7/9A | 2 | 1.78 ± 0.82 | 0.25 ± 0.11 | -0.34 ± 0.13 | 1.41 ± 0.65 | 0.11 ± 0.09 | 0.12 ± 0.04 |
| **Pallasite** | | | | | | | | |
| Zinder | UCLA 3093 | 3 | 2.01 ± 0.82 | 0.13 ± 0.11 | -0.22 ± 0.13 | 1.76 ± 0.65 | 0.06 ± 0.09 | 0.07 ± 0.04 |

Pt isotope ratios are normalized to $^{196}Pt/^{195}Pt = 0.7464$ or $^{198}Pt/^{195}Pt = 0.2145$ using the exponetial law.

The uncertainties represent the 95% confidence intervals of the mean (i.e., (s.d. × t0.95,N−1)/√N) for N ≥ 4 and for for N < 4 the 2 s.d. of the repeated analyses of the reference material NIST 129c ot the internal errors (2s.e.) of the measurements, whatever is larger.

[1] Calculated mean from Kruijer et al., 2017.

[2] Data from Worsham et al., 2019.

TABLE 2. Mo isotope data for IIIF iron meteorites, Fitwater Pass and pyroxene-bearing pallasite Zinder.

| Sample | Collection ID | N [1] | $\varepsilon^{92}Mo_{meas.}$ (± 95% CI) | $\varepsilon^{94}Mo_{meas.}$ (± 95% CI) | $\varepsilon^{95}Mo_{meas.}$ (± 95% CI) | $\varepsilon^{97}Mo_{meas.}$ (± 95% CI) | $\varepsilon^{100}Mo_{meas.}$ (± 95% CI) | $\varepsilon^{92}Mo_{CRE.corr.}$ [2] (± 95% CI) | $\varepsilon^{94}Mo_{CRE.corr.}$ [2] (± 95% CI) | $\varepsilon^{95}Mo_{CRE.corr.}$ [2] (± 95% CI) | $\varepsilon^{97}Mo_{CRE.corr.}$ [2] (± 95% CI) | $\varepsilon^{100}Mo_{CRE.corr.}$ [2] (± 95% CI) |
|---|---|---|---|---|---|---|---|---|---|---|---|---|
| **IIIF** | | | | | | | | | | | | |
| Cerro del Inca | UCLA 3092 | 8 | 1.49 ± 0.11 | 1.03 ± 0.16 | 0.86 ± 0.07 | 0.45 ± 0.04 | 0.51 ± 0.08 | 1.62 ± 0.11 | 1.12 ± 0.17 | 0.97 ± 0.08 | 0.48 ± 0.05 | 0.47 ± 0.09 |
| Clark County | BM 1959, 949 | 9 | 1.66 ± 0.22 | 1.18 ± 0.14 | 0.96 ± 0.08 | 0.49 ± 0.08 | 0.46 ± 0.08 | 1.72 ± 0.22 | 1.22 ± 0.14 | 1.00 ± 0.08 | 0.50 ± 0.08 | 0.44 ± 0.09 |
| Moonbi | UCLA S2203 | 6 | 1.55 ± 0.16 | 1.16 ± 0.10 | 0.90 ± 0.08 | 0.41 ± 0.06 | 0.51 ± 0.07 | 1.65 ± 0.16 | 1.23 ± 0.11 | 0.99 ± 0.08 | 0.43 ± 0.06 | 0.48 ± 0.07 |
| Nelson County | UCLA S2201 | 6 | 1.51 ± 0.14 | 1.11 ± 0.16 | 0.91 ± 0.06 | 0.51 ± 0.02 | 0.51 ± 0.10 | 1.55 ± 0.15 | 1.14 ± 0.16 | 0.95 ± 0.07 | 0.51 ± 0.03 | 0.50 ± 0.10 |
| St. Genevieve County | UCLA 737 | 8 | 1.44 ± 0.10 | 1.06 ± 0.10 | 0.80 ± 0.07 | 0.43 ± 0.04 | 0.56 ± 0.10 | 1.64 ± 0.11 | 1.19 ± 0.10 | 0.96 ± 0.07 | 0.47 ± 0.05 | 0.50 ± 0.10 |
| Oakley (iron) | USNM 780 | 6 | 1.39 ± 0.11 | 0.97 ± 0.11 | 0.75 ± 0.05 | 0.46 ± 0.07 | 0.49 ± 0.13 | 1.64 ± 0.12 | 1.13 ± 0.11 | 0.95 ± 0.06 | 0.51 ± 0.08 | 0.42 ± 0.13 |
| Klamath Falls [3] | ME 2789 | 8 | 1.70 ± 0.18 | 1.20 ± 0.18 | 0.98 ± 0.06 | 0.56 ± 0.11 | 0.62 ± 0.09 | 1.74 ± 0.18 | 1.22 ± 0.18 | 1.01 ± 0.07 | 0.57 ± 0.11 | 0.61 ± 0.09 |
| mean IIIF intercept [4] | | | | | | | | 1.60 ± 0.12 | 1.21 ± 0.11 | 0.98 ± 0.06 | 0.52 ± 0.03 | 0.48 ± 0.08 |
| mean IIIF mean [5] | | | | | | | | 1.63 ± 0.05 | 1.18 ± 0.05 | 0.97 ± 0.03 | 0.49 ± 0.02 | 0.47 ± 0.04 |
| Fitzwater Pass | CML 0413-7/9A | 7 | 0.16 ± 0.11 | 0.14 ± 0.09 | 0.02 ± 0.04 | 0.02 ± 0.05 | -0.03 ± 0.03 | 0.24 ± 0.11 | 0.19 ± 0.10 | 0.09 ± 0.04 | 0.03 ± 0.05 | -0.05 ± 0.03 |
| **Pallasite** | | | | | | | | | | | | |
| Zinder | UCLA 3093 | 8 | 1.12 ± 0.17 | 0.99 ± 0.12 | 0.47 ± 0.09 | 0.26 ± 0.08 | 0.21 ± 0.12 | 1.18 ± 0.17 | 1.03 ± 0.12 | 0.52 ± 0.09 | 0.27 ± 0.08 | 0.20 ± 0.12 |

The Mo isotope ratios are normalized to $^{98}Mo/^{96}Mo = 1.153173$ using the exponetial law. The reported uncertainties are the 95% confidence intervals of the mean i.e., (s.d. × t0.95,N−1)/√N) of multiple measurements of the same sample solution.

[1] Number of analyses.

[2] Mo isotope data are corrected for CRE effects using weighted mean $\varepsilon^{196}Pt$ vs. $\varepsilon^i Mo$ slopes determined for NC and CC iron meteorite groups IC, IID, IIAB, IIIAB and IIIF of −0.461±0.069 for $\varepsilon^{92}Mo$, −0.3±0.058 for $\varepsilon^{94}Mo$, −0.371±0.042 for $\varepsilon^{95}Mo$, −0.092±0.078 for $\varepsilon^{97}Mo$, and 0.130±0.057 for $\varepsilon^{100}Mo$, and are normalized to a pre-exposure $\varepsilon^{196}Pt$ of −0.06 Spitzer et al., 2021).

[3] Measured Mo data from Kruijer et al., 2017. The CRE-corrected Mo data were calculated using Pt isotope data from the same study and the $\varepsilon^{196}Pt$ vs. $\varepsilon^i Mo$ slopes determined in this study.

[4] CRE-corrected mean Mo isotope composition of IIIF iron meteorites (except Klamath Falls) derived from the intercept of the $\varepsilon^{196}Pt$ vs. $\varepsilon^i Mo$ slopes and normalized to $\varepsilon^{196}Pt = −0.06$.

[5] CRE-corrected Mo isotope composition of the IIIF iron meteorite group calculated as the weighted mean of individually CRE-corrected meteorite samples.

TABLE 3. W isotope data for Fitwater Pass and pyroxene-bearing pallasite Zinder.

| Sample | Collection ID | N | Normalized to $^{186}W/^{183}W$ | | Normalized to $^{186}W/^{184}W$ | | | $\Delta t_{CAI}$ (Ma) (± 2σ) [1] |
|---|---|---|---|---|---|---|---|---|
| | | | $\varepsilon^{182}W_{meas.}$ (± 2σ) | $\varepsilon^{184}W_{meas.}$ (± 2σ) | $\varepsilon^{182}W_{meas.}$ (± 2σ) | $\varepsilon^{183}W_{meas.}$ (± 2σ) | $\varepsilon^{182}W_{corr.}$ (± 2σ) | |
| **Iron** | | | | | | | | |
| Fitzwater Pass | CML 0413-7/9A | 6 | -2.95 ± 0.05 | -0.01 ± 0.02 | -2.92 ± 0.06 | 0.01 ± 0.03 | -2.69 ± 0.08 | 9.1 ± 1.7 |
| **Pallasite** | | | | | | | | |
| Zinder | UCLA 3093 | 6 | -3.37 ± 0.06 | 0.00 ± 0.04 | -3.35 ± 0.08 | 0.00 ± 0.05 | -3.17 ± 0.10 | 2.9 ± 1.2 |

W isotope ratios are normalized to either $^{186}W/^{183}W = 1.98590$ or $^{186}W/^{184}W = 0.92767$ using the exponetial law. The reported uncertainties are the 95% confidence intervals of the mean, i.e. (s.d. × t0.95,N−1)/√N) of multiple measurements of the same sample solution.

[1] Hf/W model age of core formation in million years relative to the formation of CAIs calculated using a $^{182}Hf$ decay constant of 0.078 Ma⁻¹, a present day chondritic $\varepsilon^{182}W$ of −1.91 ± 0.08 and a Solar System initial $\varepsilon^{182}W$ of −3.49 ± 0.07.